\documentclass[aps,pra,twocolumn, superscriptaddress, 10pt]{revtex4-2}

\usepackage{amssymb, amsmath,color,mciteplus,graphicx,subfigure}
\usepackage{calc,epsfig,epstopdf,mciteplus,bm,mathrsfs}
\usepackage{times}
\usepackage{multirow}
\usepackage{lipsum}
\usepackage[ruled,vlined,linesnumbered]{algorithm2e}
\usepackage{dcolumn}
\usepackage{setspace}
\usepackage{xcolor}
\usepackage{bm}

\begin{document}

\title{Variational Quantum Algorithm for Solving the Liouvillian Gap  }
	
	\author{Xu-Dan Xie}
	\affiliation{Key Laboratory of Atomic and Subatomic Structure and Quantum Control (Ministry of Education),  Guangdong Basic Research Center of Excellence for Structure and Fundamental Interactions of Matter, and School of Physics, South China Normal University, Guangzhou 510006, China}
	
	\author{Zheng-Yuan Xue}  \email{zyxue@scnu.edu.cn}
	\affiliation{Key Laboratory of Atomic and Subatomic Structure and Quantum Control (Ministry of Education),  Guangdong Basic Research Center of Excellence for Structure and Fundamental Interactions of Matter, and  School of Physics, South China Normal University, Guangzhou 510006, China}
	\affiliation{Guangdong Provincial Key Laboratory of Quantum Engineering and Quantum Materials,  Guangdong-Hong Kong Joint Laboratory of Quantum Matter, and Frontier Research Institute for Physics,\\ South China Normal University, Guangzhou 510006, China}
	
	\author{Dan-Bo Zhang} \email{dbzhang@m.scnu.edu.cn}
	\affiliation{Key Laboratory of Atomic and Subatomic Structure and Quantum Control (Ministry of Education),  Guangdong Basic Research Center of Excellence for Structure and Fundamental Interactions of Matter, and  School of Physics, South China Normal University, Guangzhou 510006, China}
	\affiliation{Guangdong Provincial Key Laboratory of Quantum Engineering and Quantum Materials,  Guangdong-Hong Kong Joint Laboratory of Quantum Matter, and Frontier Research Institute for Physics,\\  South China Normal University, Guangzhou 510006, China}
	
	\date{\today}

\begin{abstract}
	In open quantum systems, the Liouvillian gap characterizes the relaxation time toward the steady state. However, accurately computing this quantity is notoriously difficult due to the exponential growth of the Hilbert space and the non-Hermitian nature of the Liouvillian superoperator. In this work, we propose a variational quantum algorithm for efficiently estimating the Liouvillian gap. By utilizing the Choi–Jamiołkowski isomorphism, we reformulate the problem as finding the first excitation energy of an effective non-Hermitian Hamiltonian. Our method employs variance minimization with an orthogonality constraint to locate the first excited state and adopts a two-stage optimization scheme to enhance convergence. Moreover, to address scenarios with degenerate steady states, we introduce an iterative energy-offset scanning technique. Numerical simulations on the dissipative XXZ model confirm the accuracy and robustness of our algorithm across a range of system sizes and dissipation strengths. These results demonstrate the promise of variational quantum algorithms for simulating open quantum many-body systems on near-term quantum hardware.

\end{abstract}
	
\maketitle
\definecolor{RED}{RGB}{255,0,0}
	
\section{Introduction}
    Open quantum systems play a crucial role in a variety of fields, including quantum optics, quantum information processing, and condensed matter physics\cite{breuer2002theory,lidar2019lecture,li1986physics,rotter2015review}. Their dynamics are typically described by the Lindblad master equation, where the evolution is governed by the Liouvillian superoperator. Unlike Hermitian Hamiltonians, the Liouvillian superoperator is non-Hermitian; therefore, its spectrum is generally composed of complex eigenvalues, with the real parts dictating the relaxation rates of the system\cite{rivas2012open}. The Liouvillian gap, defined as the difference between the largest and second-largest real parts of the Liouvillian eigenvalues, governs the asymptotic relaxation toward the non-equilibrium steady state of an open quantum system, with a finite gap ensuring exponential convergence\cite{cai2013algebraic}. In contrast, the closing of the gap signals the onset of the dissipative phase transition\cite{kessler2012dissipative,minganti2018spectral,rossatto2016relaxation,li2022steady}, which are typically accompanied by critical phenomena and qualitative changes in the system's dynamical behavior. Moreover, the Liouvillian gap plays a key role in characterizing exotic non-equilibrium effects, such as the chiral damping\cite{song2019non} and the Liouvillian skin effect\cite{haga2021liouvillian,yang2022liouvillian}. As such, probing the Liouvillian gap is essential not only for uncovering the rich physics of open quantum systems but also for the practical applications in quantum sensing and computation\cite{ehrhardt2024exploring,wiersig2020robustness,verstraete2009quantum}.

Despite its importance, the direct computation of the Liouvillian gap remains a formidable challenge. Aside from a handful of exactly solvable models\cite{medvedyeva2016exact,vznidarivc2015relaxation,prosen2008third,rowlands2018noisy,shibata2019dissipative,shibata2019dissipative2,sa2020spectral,ziolkowska2020yang,nakagawa2021exact,buvca2020bethe}, the general spectral gap problem has been proven to be undecidable in terms of computational complexity\cite{lloyd1993quantum,cubitt2022undecidability,bausch2020undecidability,castilla2024undecidability}. The dimension of the Hilbert space in quantum many-body systems increases exponentially with system size, and the non-Hermitian nature of the Liouvillian superoperator exacerbates the complexity of its spectral problem. Recently, an artificial neural‑network approach has been proposed to approximate the Liouvillian gap, but it remains limited by classical resources\cite{yuan2021solving}. Meanwhile, a quantum algorithm has been suggested to tackle this problem, but it relies on the complex quantum circuit, which is difficult to implement on the current devices\cite{zhang2024exponential}.

Against this backdrop, variational quantum algorithms (VQAs) have shown promise for eigenvalue problems on near-term quantum devices\cite{moll2018quantum,bharti2022noisy,zhou2020quantum,kokail2019self,endo2021hybrid,tilly2022variational}. By minimizing the energy, they efficiently approximate ground states of Hermitian Hamiltonians\cite{wang2019accelerated,fedorov2022vqe,peruzzo2014variational,kandala2017hardware,o2016scalable,bosse2022probing} and, by enforcing orthogonality, access excited states\cite{nakanishi2019subspace,heya2023subspace,xie2022orthogonal,higgott2019variational,jones2019variational}. For non-Hermitian operators like the Liouvillian, they can also solve the steady state \cite{shang2024hermitian,yoshioka2020variational,liu2021variational}. However, the absence of orthogonality among eigenstates prevents the direct application of VQAs to compute the Liouvillian gap. A viable approach is to determine eigenvalues through variance minimization, which is applicable not only to Hermitian operators\cite{zhang2021adaptive,wang1994solving,mcclean2016theory,zhang2022variational} but also to non-Hermitian operators\cite{xie2024variational}. Nevertheless, locating the first excited state from an exponentially large space of eigenstates remains a significant challenge for solving the Liouvillian gap of the corresponding open quantum systems.

Here, we propose a VQA for estimating the Liouvillian gap of open quantum systems. Using the Choi–Jamiołkowski isomorphism\cite{zwolak2004mixed,kshetrimayum2017simple,vznidarivc2014large}, we vectorize the density matrix and reformulate the Liouvillian superoperator as a non-Hermitian Hamiltonian acting on an enlarged Hilbert space. Solving the Liouvillian gap thus maps to finding the first excited-state energy of the non-Hermitian Hamiltonian. We employ the energy variance as the cost function. To target the first excited state, we introduce a penalty term into the cost function to enforce orthogonality to the steady state, and thus remove its component. We further implement a two-stage optimization strategy, consisting of pre-training and main training, to guide the variational state toward the first excited state. To extend our method to the case with degenerate steady states, we combine the two-stage optimization strategy with an iterative energy-offset scan, gradually tuning the real part of the energy parameter to approach the Liouvillian gap.
Numerical simulations demonstrate the effectiveness of our approach in both non-degenerate and degenerate scenarios. These results highlight the promising potential of VQAs for simulating open quantum systems.

\section{master equation and the Liouvillian Gap}
   Under the Markovian approximation, an open quantum system is described by the Lindblad master equation\cite{breuer2002theory},
    \begin{equation}
        \frac{d\mathcal{\rho}}{dt}=-\text{i} [H,\rho]+\sum_j \gamma_j (L_j \rho L_j^\dagger- \frac{1}{2}\{L_j^\dagger L_j,\rho\})=:\mathcal{L}\rho,
    \end{equation}
where $[A,B]=AB-BA$ and $\{A,B\}=AB+BA$. 
Here, $\rho$ denotes the density matrix of the system and $H$ is the system's Hamiltonian, $\gamma_j$  is the dissipation rate of the $j$-th channel, and $L_j$ is the corresponding jump operator that mediates the dissipative process. Hereafter,  we assume a time‑independent Liouvillian superoperator $\mathcal{L}$, whose spectrum is determined by the eigenvalue equation, 
    \begin{equation}\label{Eq:eigen_eq}
        \mathcal{L}\rho_k=\lambda_k \rho_k,
    \end{equation}
where $\lambda_k$ is the eigenvalue and the $\rho_k$ is the corresponding eigenmatrix. As $\mathcal{L}$ is non-Hermitian, its eigenvalues are generally complex. Furthermore, it can be proven that their real parts are non‑positive\cite{breuer2002theory,rivas2012open}, and  they satisfy
    \begin{equation}
        Re(\lambda_0) \geq Re(\lambda_1) \geq Re(\lambda_2)\geq \dots,
    \end{equation}
where $\lambda_0=0$ and its corresponding eigenmatrix $\rho_0$ is the steady state of the system. Under generic conditions, the dynamics governed by $\mathcal{L}$ relax toward this steady state in the long‑time limit. Moreover, the characteristic relaxation time from an arbitrary initial state to the steady state is determined by the Liouvillian gap. When there is a unique steady state, the Liouvillian gap is  defined as
    \begin{equation}
        \Delta =|Re(\lambda_1)|.
    \end{equation}

    \section{Method}
     In this section, we propose a variational quantum algorithm to efficiently estimate the Liouvillian  gap.
     
    First, we introduce the vector represention of the master equation. we
    employ the Choi-Jamiolkowski isomorphism \cite{zwolak2004mixed,kshetrimayum2017simple}  to map the density matrix to a vector
    \begin{equation}\label{vectorize}
        \rho=\sum_{mn}\rho_{mn}|m \rangle \langle n | \Rightarrow  |\rho \rangle=\sum_{mn}\frac{\rho_{mn}}{C}|m \rangle \otimes |n \rangle 
    \end{equation} 
     where $C=\sqrt{\sum_{mn} |\rho_{mn}|^2}$ ensures normalization. Under the same mapping, the Liouvillian superoperator $\mathcal{L}$ can be represented as a non-Hermitian matrix,
    \begin{equation}
      \begin{aligned}[b]
        &\hat{\mathcal{L}} =-\text{i} H \otimes I +\text{i}I \otimes H^T  \\ 
        &+\sum_j \gamma_j (L_j  \otimes L_j^*  - \frac{1}{2} L_j^\dagger L_j  \otimes I-\frac{1}{2} I \otimes L_j^T L_j^*),
      \end{aligned}
    \end{equation}
    where $I$ is the identity matrix, $T$ denotes the matrix transpose and $*$ denotes complex conjugation. Consequently, the master equation can be written compactly as
    \begin{equation}
        \frac{d |\rho\rangle}{dt}=\hat{\mathcal{L}}|\rho\rangle.
    \end{equation}
    This mapping preserves full quantum information of $\rho$. While the density matrix $\rho$ and its vectorized form $|\rho\rangle$ may differ by a normalization factor $C$, this factor does not influence the validity of the eigenvalue equation. Specifically, when the original eigenvalue equation $\mathcal{L}\rho_k= \lambda_k \rho_k$ is mapped to its vectorized form $\hat{\mathcal{L}}|\rho_k\rangle = \lambda_k |\rho_k\rangle$, the same factor $C$ appears on both sides of the equation and thus cancels out. As a result, the eigenvalues remain invariant under this transformation, ensuring the consistency and correctness of this vectorized representation.

Then, determining the Liouvillian gap is reduced to solving the eigen equation for the first non-trivial eigenvalue $\lambda_1$ and its corresponding eigenstate $|\rho_1\rangle$, 
    \begin{equation}
        \hat{\mathcal{L}}|\rho_1 \rangle = \lambda_1 |\rho_1 \rangle.
    \end{equation}
    By analogy with closed systems, one may regard the steady state as the ground state of $\hat{\mathcal{L}}$ and $|\rho_1 \rangle$ as  its first excited state. Accordingly, the Liouvillian gap corresponds to the first excitation energy. 
    
     Due to the non-Hermitian nature of the Liouvillian superoperator, conventional VQAs based on energy minimization cannot be directly extended to compute the Liouvillian gap. Therefore, we adopt the variance as the cost function. The variance of a non-Hermitian operator $\hat{A}$ in a quantum state $|\psi\rangle$ is defined as\cite{pati2015measuring,percival1998quantum} 
            \begin{eqnarray} \label{Eq:non_var}
                \Delta A^2=\langle \psi |(\hat{A}^{\dagger}-\langle \hat{A}^{\dagger}\rangle)(\hat{A}-\langle \hat{A}\rangle)|\psi \rangle,
            \end{eqnarray}
where $\langle \hat{A}\rangle=\langle \psi |\hat{A}|\psi \rangle$ , $\langle \hat{A}^{\dagger}\rangle=\langle \psi |\hat{A}^{\dagger}|\psi \rangle$. Defining a state vector $|f\rangle = (\hat{A} -\langle\hat{A}\rangle)|\psi\rangle$, the variance can be rewritten as the  the squared norm of the state vector $|f\rangle$:
$ \Delta A^2 = \langle f|f \rangle = \| |f\rangle \|^2 \ge 0$.
In the Hilbert space, the norm of a vector is zero if and only if the vector itself is the zero vector. Therefore, we have the equivalence: $ \Delta A^2 = 0  \Leftrightarrow |f\rangle = 0$. This implies 
$\hat{A}|\psi\rangle = \langle\hat{A}\rangle|\psi\rangle$.
 Consequently, analogously to the Hermitian case, the variance vanishes if and only if the state $|\psi\rangle$ is an eigenstate of the operator $\hat{A}$  and $\langle \hat{A} \rangle$ is the corresponding eigenvalue.
 
   Based on this variational principle, the cost function can be defined as follows:
    \begin{equation}
        \mathcal{C}(E,\theta)=\langle \psi(\theta)|(\hat{\mathcal{L}}^\dagger-E^*)(\hat{\mathcal{L}}-E)|\psi(\theta)\rangle,
    \end{equation}
   where $|\psi(\theta)\rangle=U(\theta)|0\rangle$ is prepared by a parameterized unitary quantum circuit $U(\theta) $ and $E=E_r+\text{i}E_i$ represents the system's energy.
  By variationally minimizing  $\mathcal{C}(E,\theta)$ over both $\theta$ and the complex parameter 
  $E$, we can obtain the eigenvectors of $\hat{\mathcal{L}}$ and their corresponding eigenvalues $E$.

    \begin{algorithm}[tb]
      \caption{VQA for Solving the Liouvillian Gap}
      \label{alg:algorithm_1}
      \SetAlgoLined
      \begin{spacing}{1.3} 
      set initial parameters $E_{r0}=0$ \\
      update $\theta,E_i$ to minimize $\mathcal{C}_1(E_r=E_{r0},E_i,\theta)$ 
      $ \qquad \qquad  E_i',\theta' = \arg\min_{E_i, \theta}  \mathcal{C}_1(E_r=E_{r0},E_i,\theta)$ \\
      set initial parameters $E_r=E_{r0},E_i=E_i',\theta=\theta'$ \\
      update $\theta,E_i,E_r$ to minimize $\mathcal{C}(E_r,E_i,\theta)$
      $ \qquad \qquad  \tilde{E_r},\tilde{E_i} , \tilde{\theta} = \arg\min_{E_r,E_i, \theta}  \mathcal{C}(E_r,E_i,\theta)$ \\
      return $|\tilde{E_r}|$
      \end{spacing}
    \end{algorithm}
    
    However, minimizing the variance alone remains insufficient for efficiently identifying the first excited state within the exponentially large space of eigenstates. To address this, we incorporate prior knowledge about the trace structure of density matrices to constrain the search space.    
    
    Note that the eigen matrix of the Liouvillian superoperator satisfies
    \begin{equation}\label{eq:tracerho}
        \operatorname{Tr}\rho_k =\begin{cases}
         1, \quad k=0;
               \\
         0,\quad k\neq 0.
    \end{cases}
    \end{equation}
   In the vector representation, the Eq.\eqref{eq:tracerho} can be rewritten as 
   \begin{equation} \label{eq:trace_vec}
       \langle \rho_k|B\rangle=
       \begin{cases}
         c, \quad k=0;
               \\
         0,\quad k\neq 0.
        \end{cases}
   \end{equation}
    in which $c$ is a contant and $|B\rangle=\frac{1}{\sqrt{2^N}}\sum_{i=1}^{2^N} |n\rangle \otimes|n\rangle$ is the maximally entangled “Bell” state (See Appendix \ref{App:Bell} for derivation details). This implies that all excited states are orthogonal to the $|B\rangle$ in the vectorized space. 
    Therefore, We  augment our variance‑based cost function with an orthogonality penalty
    \begin{equation}
        \mathcal{C}_1(E,\theta)=\langle \psi(\theta)|(\hat{\mathcal{L}}^\dagger-E^*)(\hat{\mathcal{L}}-E)|\psi(\theta)\rangle
    +\kappa|\langle \psi(\theta)|B\rangle |^2
    \end{equation}
    where the $\kappa>0$ is a hyperparameter controlling the constraint strength. To ensure both terms have comparable scaling, $\kappa$ is typically chosen proportional to $s^2$, where $s$ is the number of Pauli strings contained in the operator $\hat{\mathcal{L}}$ (more detail seen in Appendix \ref{App:kappa}). On quantum devices, the penalty term can be obtained through Bell measurements \cite{hu2023quantum}. Minimizing the cost function $\mathcal{C}_1(\theta,E)$ both drives the variance toward zero and enforces orthogonality to the steady state, thereby ensuring convergence within the excited-state subspace.  
    
    To determine the first excited state from the exponentially large excited-state subspace and obtain the Liouvillian gap, we adopt a two-stage optimization strategy consisting of pre-training and main training, as shown in Algorithm~\ref{alg:algorithm_1}.
     For convenience, we denote $\mathcal{C}(E,\theta)$ as $\mathcal{C}(E_r,E_i,\theta)$ and $\mathcal{C}_1(E,\theta)$ as $\mathcal{C}_1(E_r,E_i,\theta)$, separating real and imaginary parts of $E$. 
     
     During the stage of pre-training, we set $E_{r0}=0$ initially. Then we keep $E_{r0}$ fixed and optimize only the parameters $\theta$ and $E_i$, thereby reducing the cost function iteratively until convergence is achieved,
     
   \begin{equation}
       E_i',\theta' = \arg\min_{E_i, \theta}  \mathcal{C}_1(E_r=E_{r0},E_i,\theta). \
   \end{equation}
   Note the first excited eigenvalue has the largest real part closest to zero.
   By holding $E_r=0$ , the variational search naturally converges toward the eigenstate whose eigenvalue has the real part close to zero, ensuring nontrivial overlap with the first excited state.
   Therefore, $|\psi(\theta')\rangle$ will achieve non-negligible overlap with the target first excited state. 
  
   In the main training stage, we set initial parameters $E_r=E_{r0},E_i=E_i',\theta=\theta'$  and then optimize all parameters to minimize the cost function $\mathcal{C}(\theta,E_r E_i)$,
   \begin{equation}
       \tilde{E_r},\tilde{E_i},\tilde{\theta} = \arg\min_{E_r, E_i, \theta}  \mathcal{C}(E_r,E_i,\theta). 
   \end{equation}
  Since the trial state  $|\psi(\theta')\rangle$ is already close to the first excited state, as the cost function (variance) converges to zero, we successfully obtain both the first excited state and the Liouvillian gap $\Delta = |\tilde{E_r}|$.

   \begin{algorithm}[tb]
      \caption{VQA for degenerate case}
      \label{alg:algorithm_2}
      \KwIn{  $m=0,\tilde{E_r}=0,\delta E$ } 
      \SetAlgoLined

      \begin{spacing}{1.3} 
      
      \While{$|\tilde{E_r}|$=0 }{
      
      set initial parameters $E_{r0}=-m*\delta E$ \\
       update $\theta,E_i$ to minimize $\mathcal{C}_1(E_r=E_{r0},E_i,\theta)$ 
      $ \qquad \qquad  E_i',\theta' = \arg\min_{E_i, \theta}  \mathcal{C}_1(E_r=E_{r0},E_i,\theta)$ \\
      set initial parameters $E_r=E_{r0},E_i=E_i',\theta=\theta'$ \\
      update $\theta,E_i,E_r$ to minimize $\mathcal{C}(E_r,E_i,\theta)$
      $ \qquad \qquad  \tilde{E_r},\tilde{E_i} , \tilde{\theta} = \arg\min_{E_r,E_i, \theta}  \mathcal{C}(E_r,E_i,\theta)$ \\
      $m=m+1$  \\
    }
    return $|\tilde{E_r}|$
      \end{spacing}
    \end{algorithm}
   The method developed thus far is restricted to systems with a unique steady state. We now generalize it to the case of degenerate steady states. Suppose $\rho_s^{(1)}$ and $\rho_s^{(2)}$ are two linearly independent steady states of $ \mathcal{L}$. Any linear combination 
   \begin{equation}
       \rho' = c_1 \rho_s^{(1)} + c_2 \rho_s^{(2)} 
   \end{equation}
   also satisfies $ \mathcal{L} \rho' = 0 $ and its trace is $\operatorname{Tr} \rho'= c_1 \operatorname{Tr}\rho_s^{(1)} + c_2 \operatorname{Tr}\rho_s^{(2)}.$ Hence, one can always choose coefficients $c_1,c_2$ such that $\operatorname{Tr} \rho'=0$, yielding a zero‑trace steady state that lies within the null space of  $\mathcal{L} $. In this case, the penalty term $ k_1 |\langle \psi(\theta) | \text{Bell} \rangle |^2 $ alone is no longer suffices to exclude all steady states from the variational search.

   To overcome this challenge, we introduce an iterative energy‑offset scan into Algorithm 1. As illustrated in Algorithm 2, we initialize the real part of the trial eigenvalue  as $ E_{r0 } = -m\cdot \delta E$, where $ \delta E $  is a small, positive increment and $m$ is an integer index. The subsequent steps follow those outlined in Algorithm \ref{alg:algorithm_1}. After each full variational optimization, if the final value of $\tilde{E}r$ remains zero, this indicates that the variational state is still confined to the steady-state manifold. In this case, the process is repeated with an increased value of $m$, thereby gradually increasing the magnitude of the initial parameter $E_{r0}$ and bringing it closer to the Liouvillian gap. This process is repeated iteratively until the nonzero $ \tilde{E_r} $ is obtained. In our numerical simulations, we typically regard $\tilde{E}_r \neq 0$ only when $\tilde{E}_r > 10^{-3}$. This signifies that $\psi(\theta)$ has exited the steady-state subspace and landed in the excited state subspace, with the optimal parameter $ |\tilde{E_r}| $ precisely equaling the Liouvillian gap. 

In practice, to choose an appropriate value for $\delta E$, we estimate the spectral range of the Liouvillian $\hat{\mathcal{L}}$ by analyzing its Hermitian part, defined as $Re(\hat{\mathcal{L}}) = (\hat{\mathcal{L}} + \hat{\mathcal{L}}^\dagger)/2$. Suppose $\text{Re}(\hat{\mathcal{L}})$ can be decomposed as $\sum_i \omega_i O_i$, where $O_i$ are Pauli strings. Since the expectation value of each $O_i$ lies within $[-1, 1]$, the operator norm of $\text{Re}(\hat{\mathcal{L}})$ is bounded by $\sum_i |\omega_i|$, i.e., $0 \leq |Re(\hat{\mathcal{L}})| < \sum_i |\omega_i|$. Assuming the eigenvalues of $\hat{\mathcal{L}}$ are approximately uniformly distributed over this spectral range, the Liouvillian gap is expected to scale as $\sum_i |\omega_i| / 2^N$, where $N$ is the number of qubits. Consequently, we typically set $\delta E < \sum_i |\omega_i| / 2^N$ to ensure sufficient resolution in the iterative energy-offset scan. This estimation can be performed efficiently and does not depend on any specific structural assumptions about $\hat{\mathcal{L}}$.

\section{result}
    \begin{figure}
        \centering
        \includegraphics[width=1\linewidth]{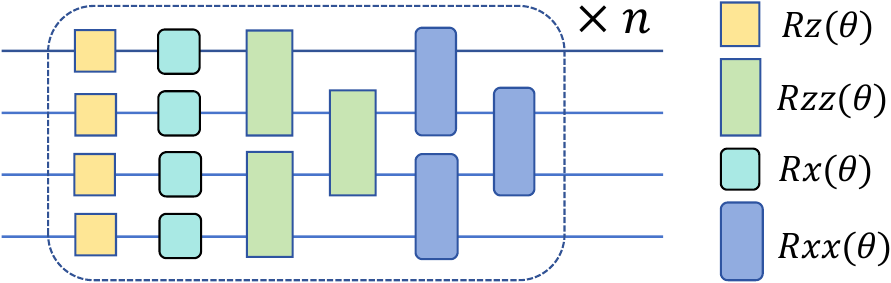}
        \caption{ Parameterized Quantum Circuit $U(\theta)$ with $n$ blocks. Here, $Rz(\theta)=exp(-\text{i} \theta \sigma^z)$, $Rx(\theta)=exp(-\text{i} \theta \sigma^x)$ ,$Rzz(\theta)=exp(-\text{i} \theta \sigma^z_j \sigma^z_{j+1})$ and $Rxx(\theta)=exp(-\text{i} \theta \sigma^x_j \sigma^x_{j+1} )$ }.
        \label{fig:circuit}
    \end{figure}
    
  In this section, we validate the effectiveness of our algorithm through numerical simulations. The VQA component is implemented with Qibo \cite{efthymiou2021qibo}, exact diagonalization is carried out using QuTiP \cite{johansson2012qutip}, and the BFGS method is adopted for optimization.
    
    To demonstrate the effectiveness and accuracy of the algorithm, we select the dissipative XXZ model as an illustrative example. In this model, the Hamiltonian is given by $H=\sum_j^{N-1} (\sigma^x_j \sigma^x_{j+1}+\sigma^y_j \sigma^y_{j+1}+J_z \sigma^z_j \sigma^z_{j+1})$ , and the jump operators are $L_j=\sigma^-_j$. Here, $\sigma^u$ denotes the Pauli operator($u=x,y,z$), and $\sigma^{\pm}=\frac{1}{2}(\sigma^x \pm \text{i} \sigma^y)$. The anisotropy parameter $J_z$ quantifies the relative strength of spin interactions along the $z$-axis compared to the $x$- and $y$-axes. 
    To compute the Liouvillian gap of this model, we employ a hardware-efficient quantum circuit. As illustrated in the fig.\ref{fig:circuit}, the circuit comprises single-qubit rotation gates and two-qubit rotation gates. 
    For the dissipative XXZ model involving $N$ spins, a total of $2N$ qubits are required for its simulation on a quantum computer. 
    
    \begin{figure}
        \centering
        \includegraphics[width=1\linewidth]{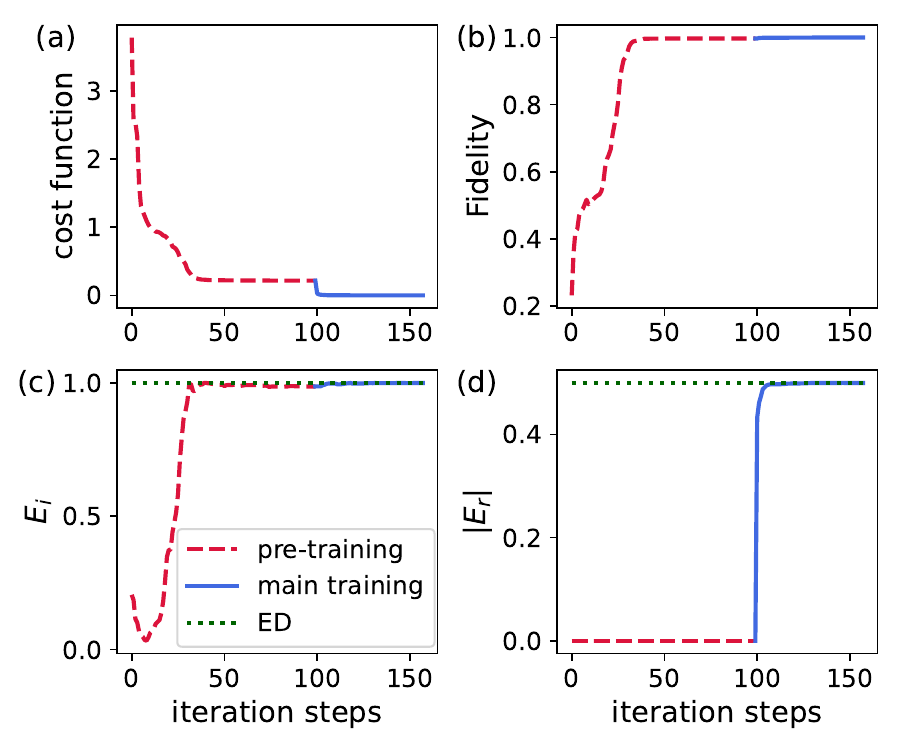}
        \caption{Numerical results for the dissipative XXZ model in this case $N=2,\gamma=1,J_z=0.5.$ (a) The variation of the cost function with the iteration steps; (b) Fidelity between $|\psi(\theta)$ and the first excited state as a function of iteration steps; (c)Evolution of the parameters $E_i$ and $E_r$ during the optimization process, as shown in (c) and (d), respectively. The red dashed line represents the pre-training process, the blue solid line indicates the main training process, and the green dotted line corresponds to the results obtained from exact diagonalization(ED).  }
        \label{fig:iteration}
    \end{figure}

    As shown in Fig.\ref{fig:iteration},  during the pre-training stage, the energy parameter $E_r$ is consistently set to zero. The cost function $\mathcal{C}_1(\theta,E)$ gradually converges to a value near zero. During the same time, the fidelity between the variational quantum state and the first excited target state steadily increases, and the energy parameter $E_i$ progressively approaches its theoretical value. Consequently, the main training stage begins with well-initialized parameters, enabling the cost function $C(\theta,E)$ to rapidly converge to zero. In parallel, the energy parameter $E_r$ exhibits remarkably fast convergence toward the theoretical value, which corresponds exactly to the Liouvillian gap.

\begin{figure}
        \centering
        \includegraphics[width=1\linewidth]{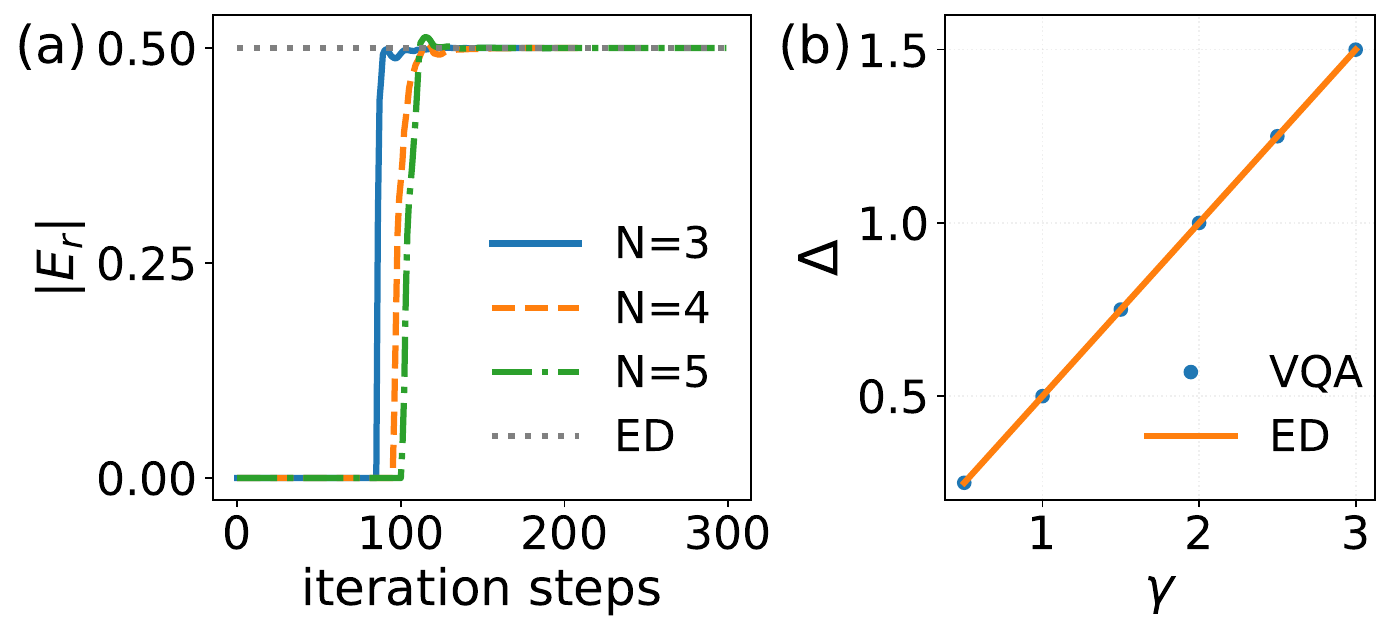}
        \caption{Comparison of the Liouvillian gap $\Delta$ obtained by VQA and the exact value. (a) $|E_r|$ as a function of iteration steps.For various system sizes $N$ with $\gamma=1$. (b) Comparison of the Liouvillian gap $\Delta$ obtained by VQA and the exact value, for various dissipation rates $\gamma$ with $N=4$. The results demonstrate that the Liouvillian gap increases linearly with $\gamma$.}
        \label{fig:gap}
    \end{figure}

    As illustrated in Fig. \ref{fig:gap}(a), we computed the Liouvillian gap of the dissipative XXZ model for different system sizes. To ensure the expressiveness of the quantum circuits, we set the block size $n$ to $2N$. The results show that the Liouvillian gap obtained via VQA closely aligns with the exact values. Notably, as the system size increases, the number of iteration steps required by the algorithm also grows. Furthermore, as depicted in Fig. \ref{fig:gap}(b), we systematically evaluated the Liouvillian gap under varying dissipation rates $\gamma$ with VQA. The numerical results exhibit excellent agreement with theoretical predictions, particularly in demonstrating a linear scaling relationship between the Liouvillian gap and the dissipation strength in the dissipative XXZ model. In summary, our algorithm performs robustly across different system sizes and dissipation rates, underscoring its effectiveness and accuracy.

    To further evaluate the performance of Algorithm \ref{alg:algorithm_2}, We consider a dissipative XXZ spin chain with pure dephasing, characterized by jump operators $L_j=\sigma^z_j$. This model exhibits a doubly degenerate steady state. To compute the gap of this model, we set $\delta E=0.3$. 

\begin{figure}
        \centering
        \includegraphics[width=1\linewidth]{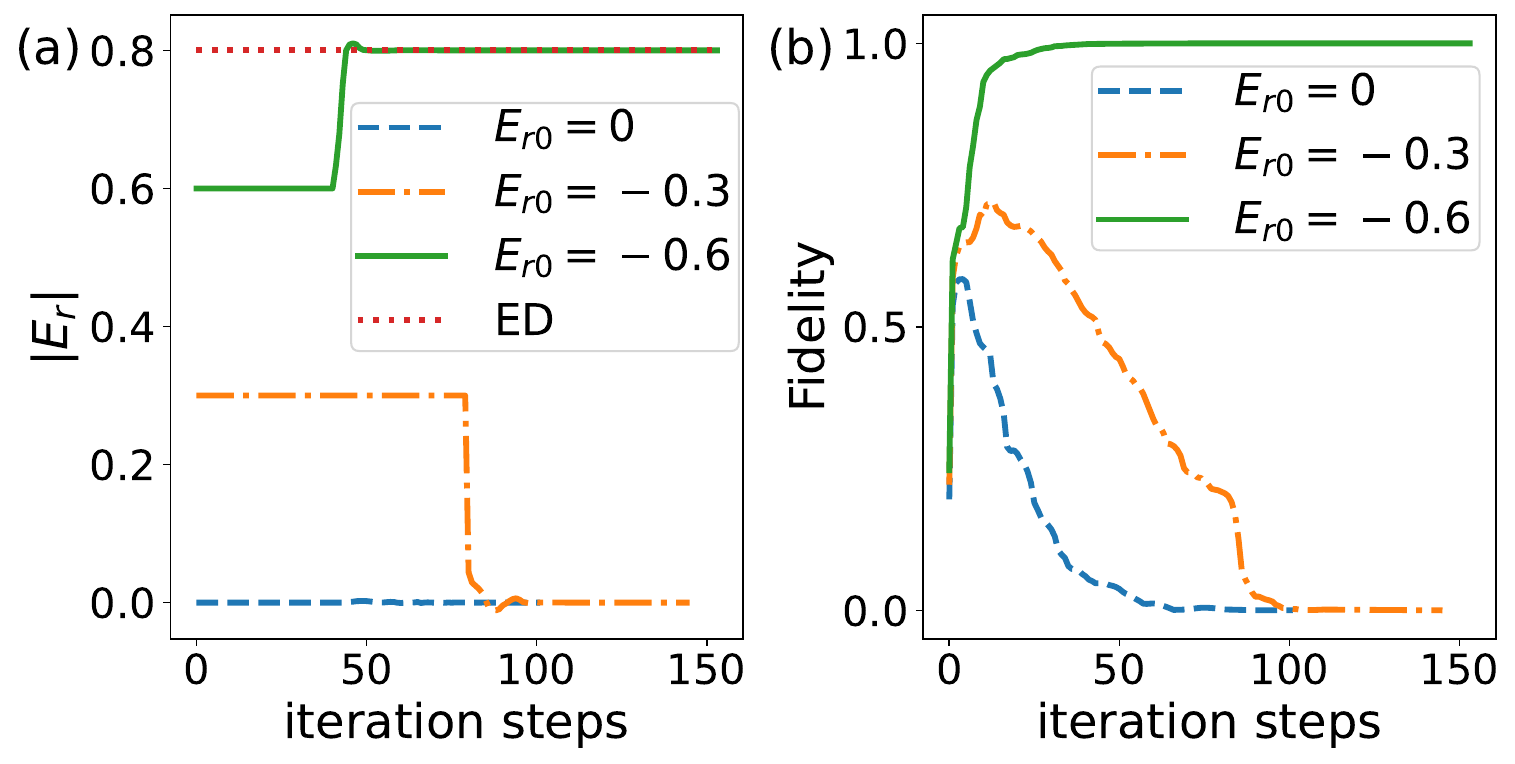}
        \caption{Numerical result for the degenerate case   in a dissipative XXZ spin chain with pure dephasing described by jump operators $L_j=\sigma^z_j$, in the case $N=2, J_z=1, \gamma=1$. (a) the variation of parameter $E_r$ during the optimization process. (b) the fidelity of $|\psi(\theta)\rangle$ and the first excited state as a function of iteration step}
        \label{fig:degen}
    \end{figure}

    As shown in Fig.\ref{fig:degen}, when initail parameter $E_{r0}=0$, the value of parameter $E_r$ remains approximately zero throughout the iteration process, significantly different from the behavior seen in Fig. \ref{fig:iteration}(d). This indicates the existence of a steady-state solution with zero trace, iconfirming that the system possesses degenerate steady states. Therefore, our algorithm can also serve as a diagnostic tool for identifying steady-state degeneracy in open quantum systems. 
    Subsequently, we adjust the initial parameter $E_{r0}=-0.3$, As shown in the results, the parameter $E_r$ still converges to zero after the iterative process, indicating that the system once again evolves toward the steady state. This suggests that when the initial guess for $E_r$ is far from the true gap, the algorithm naturally drives the system toward the steady-state subspace. We then adjust the initial parameter to $E_{r0}=-0.6$, where $E_{r0}$ becomes closer to the real part of the first excited energy. In this case, the algorithm successfully converges to the precise gap value. 
    Throughout this process, we also observe that the fidelity between the variational state 
    $|\psi(\theta)\rangle $ and the target first excited state increases monotonically. These results confirm the robustness and adaptability of our algorithm in systems with degenerate steady states.

\section{conclusion and outlook }
    In summary, we have developed a quantum algorithm for computing the Liouvillian gap that is effective for both non-degenerate and degenerate steady-state scenarios. Using the dissipative XXZ model as an illustrative example, we validated our algorithm through numerical simulations. Our work highlights the considerable potential and promising future of variational quantum algorithms in addressing the challenges of open quantum many-body systems.

    Furthermore, the development of more efficient quantum ansatz  is crucial for alleviating the barren plateau problem, which significantly hinders the trainability of variational quantum algorithms \cite{mcclean2018barren,pesah2021absence}. While some circuit designs for steady-state preparation have shown potential in avoiding barren plateaus \cite{shang2024hermitian,yoshioka2020variational}, analogous strategies for targeting excited states remain largely unexplored. Addressing this issue requires a deeper theoretical understanding of quantum circuit architectures, along with the mathematical and physical structure of Liouvillian superoperators. Moreover, the proposed variational method  can be naturally extended to classical machine learning frameworks in principle. These  will be left for future investigation.

\begin{acknowledgments}
		This work was supported by the National Natural Science Foundation of China (Grants No. 12375013 and  No.12275090), the Guangdong Basic and Applied Basic Research Fund (Grant No. 2023A1515011460),  and Guangdong Provincial Quantum Science Strategic Initiative (Grant No. GDZX2200001).
		
\end{acknowledgments}


	\appendix
         \section{\label{App:Bell} the detailed derivation of the trace of eigen matrix }
       Taking the trace of both sides of the Lindblad master equation yields
        \begin{equation}
            \mathrm{Tr}(\mathcal{L}\rho)=-i\mathrm{Tr}([H,\rho])+\sum_{i}\gamma_i\mathrm{Tr}(L_{i}\rho L_{i}^{\dagger}-\frac{1}{2}\{L_{i}^{\dagger}L_{i},\rho\}).
        \end{equation}
       Because
        \begin{equation}
            \mathrm{Tr}([H,\rho])=\mathrm{Tr}(H\rho)-\mathrm{Tr}(\rho H)=0
        \end{equation}
        and
        \begin{eqnarray}
           &&\mathrm{Tr}(L_{i}\rho L_{i}^{\dagger}-\frac{1}{2}\{L_{i}^{\dagger}L_{i},\rho\}) \\ \notag
            &&=\mathrm{Tr}(L_{i}\rho L_{i}^{\dagger})-\frac{1}{2}\mathrm{Tr}(L_{i}^{\dagger}L_{i}\rho+\rho L_{i}^{\dagger}L_{i}) \\ \notag
            &&=\mathrm{Tr}(L_{i}^{\dagger}L_{i}\rho) -\mathrm{Tr}(L_{i}^{\dagger}L_{i}\rho) =0,
        \end{eqnarray}
            
      we get  $\mathrm{Tr}(\mathcal{L}\rho)=0$.Therefore,
      \begin{eqnarray}
          \mathrm{Tr}(\mathcal{L}\rho_k)=\lambda_k \mathrm{Tr}(\rho_k)=0
      \end{eqnarray}
      So the trace of the eigen matrix of the Liouvillian superoperator satisfies
      \begin{equation}\label{eq:trace_A}
         \mathrm{Tr}(\rho_k)=\begin{cases}
         1, \quad k=0;
               \\
         0,\quad k\neq 0.
    \end{cases}
    \end{equation}
    In the repretation of vector, $|\rho \rangle=\sum_{mn}\frac{\rho_{mn}}{C}|m \rangle \otimes |n \rangle $. So
    \begin{eqnarray}
        \big(\langle n|\otimes \langle n| \big)|\rho\rangle = \frac{\rho_{nn}}{C}.
    \end{eqnarray}
    Hence,
    \begin{eqnarray}
         \mathrm{Tr}(\rho)&&=\sum_n \rho_{nn}=C\sum_n \big(\langle n|\otimes \langle n|  \big)|\rho\rangle \\ \notag
         &&=C \sqrt{2^N} \langle Bell| \rho \rangle,
    \end{eqnarray}
   where $|B\rangle=\frac{1}{\sqrt{2^N}}\sum_{i=1}^{2^N} |n\rangle \otimes|n\rangle$. According to Eq.~\ref{eq:trace_A}, we get
   \begin{equation} \label{eq:trace_v_A}
       \langle B| \rho_k\rangle=
       \begin{cases}
         c, \quad k=0;
               \\
         0,\quad k\neq 0.
        \end{cases}
   \end{equation}

\section{\label{App:kappa} the choice of the hyperparameter $\kappa$ }
   Here, we discuss the choice of the hyperparameter $\kappa$. The cost function $\mathcal{C}_1$ consists of two terms. The first term is a variance term, given by

\begin{eqnarray}
    \langle \psi(\theta) | (\hat{\mathcal{L}}^\dagger - E^*)(\hat{\mathcal{L}} - E) | \psi(\theta) \rangle = \langle \psi(\theta) | \hat{M} | \psi(\theta) \rangle,
\end{eqnarray}
where $\hat{M} = (\hat{\mathcal{L}}^\dagger - E^*)(\hat{\mathcal{L}} - E).$

Suppose $\mathcal{L} = \sum_{i=1}^s \omega_i \hat{O}_i,$ where each $\hat{O}_i$ is a Pauli string, and there are $s$ such terms in total. Since $\hat{M}$ is approximately the Hermitian conjugate of $\mathcal{L}$ multiplied by itself, its expansion generally contains on the order of $\mathcal{O}(s^2)$ Pauli strings.
For a variational quantum state $|\psi(\theta)\rangle$, the expectation value of each individual Pauli string $\langle \psi(\theta) | \hat{O}_i | \psi(\theta) \rangle$ typically scales as $\mathcal{O}(1)$. Consequently, the total expectation
$\langle \psi(\theta) | \hat{M} | \psi(\theta) \rangle$ can be estimated to scale as $\mathcal{O}(s^2)$. 

The second term of the cost function, $|\langle \psi(\theta) | \mathrm{Bell} \rangle|^2,$
represents the squared overlap between the variational state and the Bell state, and also scales as $\mathcal{O}(1)$.
To ensure both terms contribute comparably to the cost function, the hyperparameter $\kappa$ should be chosen to scale as $\mathcal{O}(s^2)$. Therefore, for dissipative XXZ model with sistem size $N$, we choose $\kappa=N^2$. 

Alternatively, the cost function $\mathcal{C}_1$ can be understood from the viewpoint of multi-objective optimization. One may also consider tackling this problem using techniques from constrained optimization, such as Lagrange multiplier method \cite{bertsekas2014constrained} and Pareto optimization \cite{cheng2025variational,lin2019pareto}.

\bibliographystyle{apsrev4-2}
\bibliography{ref}

	
\end{document}